\documentstyle[twoside, psfig]{article}

\catcode`\@=11
\long\def\@makefntext#1{
\protect\noindent \hbox to 3.2pt {\hskip-.9pt  
$^{{\eightrm\@thefnmark}}$\hfil}#1\hfill}		

\def\thefootnote{\fnsymbol{footnote}}
\def\@makefnmark{\hbox to 0pt{$^{\@thefnmark}$\hss}}	
	
\def\ps@myheadings{\let\@mkboth\@gobbletwo
\def\@oddhead{\hbox{}
\rightmark\hfil\eightrm\thepage}   
\def\@oddfoot{}\def\@evenhead{\eightrm\thepage\hfil
\leftmark\hbox{}}\def\@evenfoot{}
\def\sectionmark##1{}\def\subsectionmark##1{}}

\oddsidemargin=\evensidemargin
\renewcommand{\thefootnote}{\fnsymbol{footnote}}

\newcounter{sectionc}\newcounter{subsectionc}\newcounter{subsubsectionc}
\renewcommand{\section}[1] {\vspace{12pt}\addtocounter{sectionc}{1} 
\setcounter{subsectionc}{0}\setcounter{subsubsectionc}{0}\noindent 
	{\tenbf\thesectionc. #1}\par\vspace{5pt}}
\renewcommand{\subsection}[1] {\vspace{12pt}\addtocounter{subsectionc}{1} 
	\setcounter{subsubsectionc}{0}\noindent 
	{\bf\thesectionc.\thesubsectionc. {\kern1pt \bfit #1}}\par\vspace{5pt}}
\renewcommand{\subsubsection}[1] {\vspace{12pt}\addtocounter{subsubsectionc}{1}
	\noindent{\tenrm\thesectionc.\thesubsectionc.\thesubsubsectionc.
	{\kern1pt \tenit #1}}\par\vspace{5pt}}

\newcounter{appendixc}
\newcounter{subappendixc}[appendixc]
\newcounter{subsubappendixc}[subappendixc]
\renewcommand{\thesubappendixc}{\Alph{appendixc}.\arabic{subappendixc}}
\renewcommand{\thesubsubappendixc}
	{\Alph{appendixc}.\arabic{subappendixc}.\arabic{subsubappendixc}}

\renewcommand{\appendix}[1] {\vspace{12pt}
        \refstepcounter{appendixc}
        \setcounter{figure}{0}
        \setcounter{table}{0}
        \setcounter{lemma}{0}
        \setcounter{theorem}{0}
        \setcounter{corollary}{0}
        \setcounter{definition}{0}
        \setcounter{equation}{0}
        \renewcommand{\thefigure}{\Alph{appendixc}.\arabic{figure}}
        \renewcommand{\thetable}{\Alph{appendixc}.\arabic{table}}
        \renewcommand{\theappendixc}{\Alph{appendixc}}
        \renewcommand{\thelemma}{\Alph{appendixc}.\arabic{lemma}}
        \renewcommand{\thetheorem}{\Alph{appendixc}.\arabic{theorem}}
        \renewcommand{\thedefinition}{\Alph{appendixc}.\arabic{definition}}
        \renewcommand{\thecorollary}{\Alph{appendixc}.\arabic{corollary}}
        \renewcommand{\theequation}{\Alph{appendixc}.\arabic{equation}}
        \noindent{\tenbf Appendix \theappendixc #1}\par\vspace{5pt}}
\newcommand{\subappendix}[1] {\vspace{12pt}
        \refstepcounter{subappendixc}
        \noindent{\bf Appendix \thesubappendixc. {\kern1pt \bfit #1}}
	\par\vspace{5pt}}
\newcommand{\subsubappendix}[1] {\vspace{12pt}
        \refstepcounter{subsubappendixc}
        \noindent{\rm Appendix \thesubsubappendixc. {\kern1pt \tenit #1}}
	\par\vspace{5pt}}

\newcommand{\textlineskip}{\baselineskip=13pt}
\newcommand{\smalllineskip}{\baselineskip=10pt}

\def\eightcirc{
\begin{picture}(0,0)
\put(4.4,1.8){\circle{6.5}}
\end{picture}}
\def\eightcopyright{\eightcirc\kern2.7pt\hbox{\eightrm c}}

\def\abstracts#1#2#3{{
	\centering{\begin{minipage}{4.5in}\baselineskip=10pt\footnotesize
	\parindent=0pt #1\par 
	\parindent=15pt #2\par
	\parindent=15pt #3
	\end{minipage}}\par}} 

\renewenvironment{thebibliography}[1]
	{\frenchspacing
	 \ninerm\baselineskip=11pt
	 \begin{list}{\arabic{enumi}.}
	{\usecounter{enumi}\setlength{\parsep}{0pt}
	 \setlength{\leftmargin 12.7pt}{\rightmargin 0pt} 
	 \setlength{\itemsep}{0pt} \settowidth
	{\labelwidth}{#1.}\sloppy}}{\end{list}}

\newcounter{itemlistc}
\newcounter{romanlistc}
\newcounter{alphlistc}
\newcounter{arabiclistc}

\newcommand{\fcaption}[1]{
        \refstepcounter{figure}
        \setbox\@tempboxa = \hbox{\footnotesize Fig.~\thefigure. #1}
        \ifdim \wd\@tempboxa > 5in
           {\begin{center}
        \parbox{5in}{\footnotesize\smalllineskip Fig.~\thefigure. #1}
            \end{center}}
        \else
             {\begin{center}
             {\footnotesize Fig.~\thefigure. #1}
              \end{center}}
        \fi}

\newcommand{\tcaption}[1]{
        \refstepcounter{table}
        \setbox\@tempboxa = \hbox{\footnotesize Table~\thetable. #1}
        \ifdim \wd\@tempboxa > 5in
           {\begin{center}
        \parbox{5in}{\footnotesize\smalllineskip Table~\thetable. #1}
            \end{center}}
        \else
             {\begin{center}
             {\footnotesize Table~\thetable. #1}
              \end{center}}
        \fi}

\def\@citex[#1]#2{\if@filesw\immediate\write\@auxout
	{\string\citation{#2}}\fi
\def\@citea{}\@cite{\@for\@citeb:=#2\do
	{\@citea\def\@citea{,}\@ifundefined
	{b@\@citeb}{{\bf ?}\@warning
	{Citation `\@citeb' on page \thepage \space undefined}}
	{\csname b@\@citeb\endcsname}}}{#1}}

\newif\if@cghi
\def\cite{\@cghitrue\@ifnextchar [{\@tempswatrue
	\@citex}{\@tempswafalse\@citex[]}}
\def\citelow{\@cghifalse\@ifnextchar [{\@tempswatrue
	\@citex}{\@tempswafalse\@citex[]}}
\def\@cite#1#2{{$\null^{#1}$\if@tempswa\typeout
	{IJCGA warning: optional citation argument 
	ignored: `#2'} \fi}}

\def\pmb#1{\setbox0=\hbox{#1}
	\kern-.025em\copy0\kern-\wd0
	\kern.05em\copy0\kern-\wd0
	\kern-.025em\raise.0433em\box0}

\def\fnt#1#2{\footnotetext{\kern-.3em
	{$^{\mbox{\scriptsize #1}}$}{#2}}}

\def\fpage#1{\begingroup
\voffset=.3in
\thispagestyle{empty}\begin{table}[b]\centerline{\footnotesize #1}
	\end{table}\endgroup}

\def\runninghead#1#2{\pagestyle{myheadings}
\markboth{{\protect\footnotesize\it{\quad #1}}\hfill}
{\hfill{\protect\footnotesize\it{#2\quad}}}}
\font\tenrm=cmr10
\font\tenit=cmti10 
\font\tenbf=cmbx10
\font\bfit=cmbxti10 at 10pt
\font\ninerm=cmr9

\font\eightrm=cmr8

\def\qed{\hbox{${\vcenter{\vbox{			
   \hrule height 0.4pt\hbox{\vrule width 0.4pt height 6pt
   \kern5pt\vrule width 0.4pt}\hrule height 0.4pt}}}$}}

\renewcommand{\thefootnote}{\fnsymbol{footnote}}	

\begin{document}

\runninghead{ Dynamical Supersymmetry Breaking $\ldots$} 
{ Dynamical Supersymmetry Breaking  $\ldots$}

\normalsize\textlineskip
\thispagestyle{empty}
\setcounter{page}{1}


hep-ph/9710274 \hfill UCSD/PTH-97-30

September 1997

\vspace*{0.88truein}

\fpage{1}
\centerline{\bf DYNAMICAL SUPERSYMMETRY BREAKING---}
\vspace*{0.035truein}
\centerline{\bf WHY AND HOW\footnote{Based on a talk given at the SLAC Experimental
Seminar, April 21, 1997.}}
\vspace*{0.37truein}
\centerline{ \footnotesize ERICH POPPITZ \footnote{poppitz@einstein.ucsd.edu} }
\vspace*{0.015truein}
\centerline{\footnotesize\it Department of Physics, University
of California at San Diego}
\baselineskip=10pt
\centerline{\footnotesize\it La Jolla, CA 92093-0319, U S A }
\vspace*{10pt}

\vspace*{0.225truein}

\vspace*{0.21truein}
\abstracts{
This theoretical review is intended to give
non-theorists a flavor of the ideas driving 
the current efforts to experimentally find 
supersymmetry. We discuss the main reasons  
behind the  expectation that supersymmetry may be 
``just around the corner" and may be discovered in the near future.
We use simple quantum-mechanical examples to illustrate the 
concept---and the power---of supersymmetry, the possible ways 
to break supersymmetry, and the dynamical generation of small 
scales. We then describe how this theoretical machinery helps 
shape our perception of what  physics beyond 
the electroweak scale might be.}{}{}


\vspace*{1pt}\textlineskip	

\noindent


\textheight=7.8truein
\setcounter{footnote}{0}
\renewcommand{\thefootnote}{\alph{footnote}}

\section{Introduction}

Since the invention of supersymmetry
more than 25 years ago,  physicists have been 
fascinated\footnote{As of the date of submission of this article, 
8447 publications on supersymmetry---both 
theoretical and experimental---were 
listed in the SLAC SPIRES-HEP database.} \hspace{.1cm} by the possibility 
that this new fundamental space-time symmetry might govern
physics at short distances. 
Theorists have 
devoted their research 
to both the mathematical aspects and the applications
of supersymmetry to elementary particle physics. 
At the same time, experimentalists 
(not without the encouragement of their theoretical colleagues) have been 
continually searching for supersymmetry at increasingly higher  
energy scales, with enduring negative results. Will this 
process of continually probing new energy scales, looking for 
supersymmetry, and not finding it (and consequently setting new, higher 
benchmarks for its discovery) continue indefinitely? 
At which point are we willing to give up the idea
that supersymmetry has anything to do with experimentally 
accessible particle physics? We believe that currently
 the search for supersymmetry has reached an important 
threshold. Within the next 10 years---with the advent of the 
Large Hadron Collider---we will have 
the answer to the question: ``Is supersymmetry relevant for physics 
at the electroweak scale?"
 
In this article, we review the main 
 properties of supersymmetry
that make it an attractive possibility for physics beyond the 
standard model (Sections 2 and 3). We explain why we believe that 
if supersymmetry is relevant for electroweak scale physics, it must
be {\it dynamically} broken. Some quantum-mechanical examples
 illustrating  how supersymmetry can  
break are discussed in Section 4. We point out 
 that  the central theoretical problem of extending what 
 has become known as the Minimal 
Supersymmetric Standard Model (MSSM) 
 to a consistent theoretical framework is 
the mechanism of  supersymmetry breaking.
We describe the current theoretical ideas of how supersymmetry breaks in Section 5. 
We conclude by  stressing
 the importance of experimental input in sharpening these ideas:
 if supersymmetry is found 
and  the spectrum of superparticles measured, we will get clues
 pointing towards the likely mechanism of supersymmetry breaking.


\noindent 

\section{What is supersymmetry?}

{\flushleft{\it Supersymmetry 
is a new space-time symmetry interchanging bosons and fermions}}. In order 
to clarify this concise definition, we will introduce supersymmetry via the 
simplest supersymmetric quantum-mechanical 
system, the ``supersymmetric oscillator." The supersymmetric oscillator
 is a simple generalization of the one dimensional harmonic oscillator.
In this section, we will discuss its properties in some detail 
(as an aside, we will see that the supersymmetric oscillator describes a  
well-known physical system: that of an electron moving in a constant, 
homogeneous magnetic field). 
Many of the properties of the supersymmetric oscillator do, in fact, 
generalize to quantum field theory and are behind the reasons 
that make supersymmetry an attractive possibility for physics beyond 
the electroweak scale. We will end this section by concluding that supersymmetry,
if it is relevant for  elementary particle physics, 
can not be realized in its simplest form---as 
it is in the supersymmetric oscillator---but rather has to be broken at
an energy scale at order or above the electroweak scale.

Supersymmetry is a symmetry that
relates bosons and fermions. The simplest quantum-mechanical 
system that leads
to the introduction of bosons is the harmonic oscillator. Recall that the
harmonic oscillator  describes the motion of a particle in one spatial 
dimension (with coordinate denoted by $x$) in a quadratic potential 
$V(x) = \omega^2 x^2/2$. The stationary 
states of the particle in the harmonic well 
is described by its wave function, $\Psi(x)$, 
which obeys the Schr\" odinger equation $\hat{H} \Psi(x) = E \Psi(x)$.
The Hamiltonian has the form  (we 
use units where the mass of the particle equals one):
\begin{equation}
\label{HB} 
\hat{H}_B ~=~ - \frac{\hbar^2}{2} 
  \frac{d^2}{d x^2} ~+ ~V(x)~=~
  \hbar~ \omega~\left( b^\dagger ~b + {1\over 2} \right)~.
\end{equation}
The second equality follows
from the substitution
$b = \sqrt{\hbar \over 2 \omega} {d\over dx} + \sqrt{\omega\over 2 \hbar} x$, 
$b^\dagger = - \sqrt{\hbar \over 2 \omega} {d\over dx} + \sqrt{\omega\over 2 \hbar} x$.
The commutation relation
 $[{d\over d x}, x] = 1$ implies that the operators $b^\dagger, b$ obey 
 the {\it commutation} relation:
 \begin{equation}
 \label{bosecommutator}
 \left[ ~b, b^\dagger \right] ~=~ b~b^\dagger - ~b^\dagger ~b~=~1~.
 \end{equation}
 The spectrum of  allowed energy levels is labeled by a single quantum
 number, $n = 0, 1, 2,...$, and is 
 given by the well-known formula:
 \begin{equation}
 \label{boselevels}
 E_n^B ~=~\hbar ~\omega \left( n + {1\over 2} \right)~, n = 0, 1, 2, ...~.
\end{equation}
The spectrum (\ref{boselevels}) of the bosonic oscillator is shown on Fig.~1. 

\begin{figure}[ht]
\vspace*{13pt}
\centerline{\vbox{\hrule width 5cm height0.001pt}}
\vspace*{.1truein}
\centerline{ \psfig{file=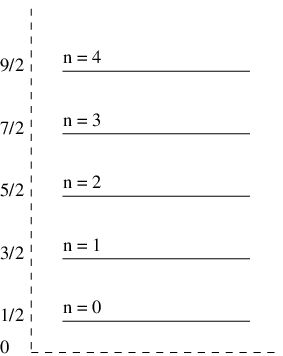}}
\centerline{\vbox{\hrule width 5cm height0.001pt}}
\vspace*{13pt}
\fcaption{The energy levels of the bosonic oscillator (in units of $\hbar \omega$).}
\end{figure}

The commutation relation (\ref{bosecommutator}) is typical for systems of 
bosons; the operators $b^\dagger$ and $b$ are called bosonic creation and
annihilation operators, respectively. 
One can interpret the ground state  state $\vert 0 \rangle$ 
of the harmonic oscillator as a state with no quanta, the state
$\vert 1 \rangle \sim b^\dagger \vert 0 \rangle$ as a state with one
quantum, and, generally, 
the state $\vert n \rangle \sim (b^\dagger)^n \vert 0 \rangle$ 
as a state with $n$ quanta. The operator $b^\dagger$ is called a 
bosonic creation operator, because its action of a state with $n$ quanta
 creates an additional quantum, i.e. a state with $n + 1$ quanta. Similarly,
 the operator $b$ decreases the occupation number and annihilates a quantum 
 from the state upon which it acts. 
 
 The interpretation of the harmonic oscillator in this
  ``second quantized" representation 
  is very useful when describing the quantization of, e.g.,
  the Maxwell field. Later in this section, 
  we will use this interpretation to generalize
  the supersymmetric oscillator to quantum field theory.

Now we can go on, and, by analogy with the bosonic oscillator, introduce
what can be called the ``fermionic oscillator." The definition below
 may initially seem somewhat formal, but later in this section we will 
 discuss a physical example. It will become clear
that the fermionic oscillator describes many quantum mechanical physical systems, 
in particular systems with two energy levels (spin up/spin down).
The bosonic oscillator can be formally
 defined as the system whose Hamiltonian
operator 
is given by (\ref{HB}), with the bosonic creation and
annihilation operators obeying the commutation relation (\ref{bosecommutator}). 
Similarly
one can define the fermionic oscillator by its Hamiltonian:
\begin{equation}
\label{HF}
\hat{H}_F ~=~ \hbar~\omega ~\left( f^\dagger ~f - {1\over 2}\right)~,
\end{equation}
where the fermion creation ($f^\dagger$) and annihilation ($f$)
 operators obey the {\it anticommutation} relation
 \begin{equation}
 \label{fermicommutator}
 \left\{ f, f^\dagger \right\} ~=~ f ~f^\dagger ~+ ~f^\dagger~ f~= ~1~,
 \end{equation}
as well as 
\begin{equation}
\label{nilpotence}
f^2 ~=~ (f^\dagger)^2 ~=~ 0~.
\end{equation} 
The latter property (called nilpotence) should
be reminiscent of the Pauli principle---no two fermions can occupy the 
same quantum state. One can be even more explicit and use the following 
representation for the fermionic creation and
annihilation operators  by two by two matrices:
\begin{eqnarray}
\label{fermionrep}f &=& \left(  \begin{array}{cc}    0 & 0 \\ 
                                    1 & 0  \end{array} \right)  \nonumber \\
 f^\dagger &=& \left( \begin{array}{cc} 
                              0 & 1 \\
                              0 & 0 
\end{array}
 \right)  
~.
\end{eqnarray}
By matrix manipulation, it is easy to see that the operators (\ref{fermionrep})
obey (\ref{fermicommutator}) 
and (\ref{nilpotence}).
The representation (\ref{fermionrep}) also makes it easy to see that 
the Hamiltonian of the fermionic oscillator, (\ref{HF}), describes a two-level
system with energy levels 
\begin{equation}
\label{fermilevels}
E_k^F ~=~ \hbar~\omega ~\left( k - {1\over 2} \right)~, k = 0, 1~.
\end{equation}
The spectrum (\ref{fermilevels}) of the fermionic oscillator is shown on Fig.~2.

\begin{figure}[ht]
\vspace*{13pt}
\centerline{\vbox{\hrule width 5cm height0.001pt}}
\vspace*{.1truein}	
\centerline{\psfig{file=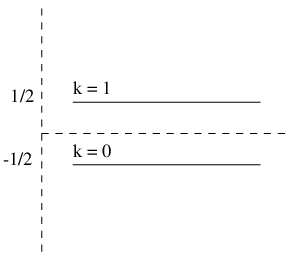}}
\centerline{\vbox{\hrule width 5cm height0.001pt}}
\vspace*{13pt}
\fcaption{The energy levels of the fermionic oscillator (in units of $\hbar \omega$).}
\end{figure}

The operators $f^\dagger$ and $f$ can be interpreted as creation and annihilation
operators, similar to the interpretation given to $b^\dagger$ and $b$ 
in the bosonic oscillator. The difference 
is that the anticommuting nature of 
$f^\dagger, f$, and their nilpotence (\ref{nilpotence}) imply
that it is impossible to create two fermion quanta in the same state by 
applying $f^\dagger$ twice. Thus  (\ref{fermicommutator}) and
(\ref{nilpotence}) ensure that  the fermionic 
oscillator obeys the Pauli 
principle. Correspondingly, the quantum number $k = 0,1$ can be 
 called the fermion 
occupation number.

We are now ready to define the supersymmetric oscillator. 
It is simply the sum of
the bosonic and fermionic oscillators with equal
 spacing $\hbar \omega$ of energy levels, zero-point energy $\hbar \omega/2$ 
 for the bosonic, and $- \hbar \omega/2$ for the fermionic oscillator (for any other 
 choice of the fermion zero point energy, 
 the resulting oscillator will not be supersymmetric; see the discussion below). 
 The Hamiltonian of the supersymmetric oscillator
will therefore be:
\begin{equation}
\label{HSUSY}
\hat{H}_{SUSY} ~= ~\hat{H}_F~+~\hat{H}_B ~=~ 
\hbar~\omega~\left( ~b^\dagger ~b ~+~ f^\dagger ~f ~\right) ~.
\end{equation}
The bosonic and fermionic creation and annihilation operators obey the 
commutation (\ref{bosecommutator}) and anticommutation (\ref{fermicommutator})
relations, respectively  (while all
bosonic operators commute with all fermionic operators), and the 
fermionic operators obey (\ref{nilpotence}).
The Hamiltonian (\ref{HSUSY}) can  be written in several equivalent 
forms using the (anti-)commutation relation (\ref{fermicommutator}) 
and (\ref{nilpotence}):
\begin{eqnarray}
\label{HSUSY1}
\hat{H}_{SUSY} ~&=&~\hbar~\omega~\left( ~b^\dagger ~b ~+
~ f^\dagger ~f ~\right) \nonumber \\
~&=&~ \hbar~\omega~\left(~ b^\dagger ~f ~+~ f^\dagger ~b~ \right)^2~ 
\nonumber \\
&=&~\hbar~\omega~\left(~ Q^\dagger ~+~ Q ~ \right)^2  \\
&=&~\hbar~\omega~\left\{ ~Q^\dagger, Q ~\right\} \nonumber~.
\end{eqnarray}
Here we have defined the operators 
\begin{equation}
\label{SUSYgenerators}
Q ~=~b^\dagger~ f, ~~~ Q^\dagger ~=~f^\dagger~b ~~.
\end{equation}
These operators are called, for reasons to become clear below, {\it 
supersymmetry
generators}.
It is easy to see from (\ref{SUSYgenerators}) 
 that they obey the same kind of property (\ref{nilpotence}) (nilpotence) as
$f^\dagger$ and $f$: $Q^2 = (Q^\dagger)^2 = 0$ (this was essential
 for going from the 
third to the last line in (\ref{HSUSY1})). 
The  generators of supersymmetry
 $Q, Q^\dagger$ can be easily seen, using (\ref{HSUSY1}) and (\ref{nilpotence}),
  to commute with the Hamiltonian:
\begin{equation}
\label{susyH}
\left[ Q, H_{SUSY} \right] ~=~ \left[ Q^\dagger, H_{SUSY} \right] ~=~ 0~.
\end{equation}
By absorbing the factor $\hbar \omega $ in a redefinition of the Hamiltonian,
 their anticommutation relation can be written as 
\begin{equation}
\label{susyGENS}
\left\{ Q^\dagger, Q \right\} ~= ~ \hat{H}~.
\end{equation}

The energy spectrum of $H_{SUSY}$ is simply the sum of the 
energy spectra of the bosonic (Fig.~1) and fermionic (Fig.~2)
 oscillators. The energy levels are labeled  by the  two quantum 
numbers, $(n,k)$, with $n = 0,1,2,...$ and $k = 0,1$ (the 
bosonic and fermionic occupation
 numbers inherited from the 
bosonic and fermionic oscillator):
 \begin{equation}
 \label{SUSYSPECTRUM}
 E_{n,k}^{SUSY} ~=~ \hbar ~\omega~\left( n + k \right)~.
 \end{equation}
 The spectrum  (\ref{SUSYSPECTRUM}) of the supersymmetric oscillator is shown on
 Fig.~3.

\begin{figure}[ht]
\vspace*{13pt}
\centerline{\vbox{\hrule width 5cm height0.001pt}}
\vspace*{.1truein}	
\centerline{\psfig{file=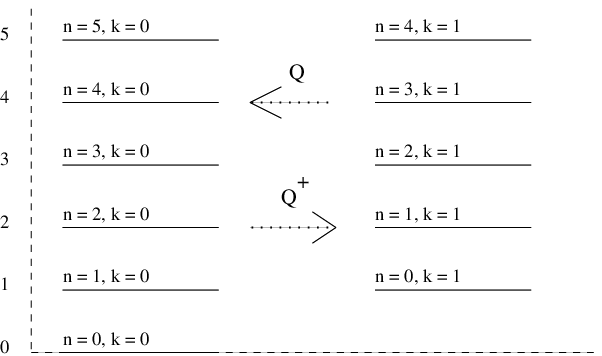}}
\centerline{\vbox{\hrule width 5cm height0.001pt}}
\vspace*{13pt}
\fcaption{The energy levels of the supersymmetric
 oscillator (in units of $\hbar \omega$). The action of the supersymmetry generators
 $Q$ and $Q^\dagger$ on the degenerate 
energy levels (forming a supermulitplet) 
is shown by the corresponding 
 arrows.}
\end{figure} 

All energy levels in the spectrum are doubly degenerate---an energy
level with $n+1$ bosons and no fermions is degenerate with the level
with $n$ bosons and one fermion. This Bose-Fermi
degeneracy is due to supersymmetry: we noted above
that the operators $Q$ and $Q^\dagger$ (\ref{susyGENS}) 
commute with the Hamiltonian (\ref{susyH}). As usual, the existence of operators
that commute with the Hamiltonian indicates that the energy levels are degenerate.
The novelty in the case of supersymmetry is that the operators that generate the 
symmetry (i.e.  commute with the Hamiltonian)  themselves obey
{\it anticommutation}
(rather then the usual commutation)
relations and thus relate bosons to fermions: we see that
 the operator $Q^\dagger = f^\dagger b$ (which creates a fermion and
annihilates a boson) brings us from  an energy level
labeled by the bosonic and fermionic occupation numbers 
$(n +1, 0)$ to the $(n, 1)$ level; similarly, the operator 
$Q = b^\dagger f$ creates a boson and annihilates a fermion 
and relates the $(n,1)$ level to the $(n+1,0)$ 
level. The degenerate levels related by the action of the supersymmetry generators
are said to form {\it supermultiplets}.

In addition to the degeneracy of the energy levels of bosons and fermions,
supersymmetric systems have  other general properties that are
important in   applications of supersymmetry to elementary particle physics.
Since the Hamiltonian is a total square (see the 
second equation in (\ref{HSUSY1})) of a hermitean operator, 
all its   states have nonnegative energies.
In particular, the ground state (if supersymmetry is unbroken) always has 
vanishing  energy (see Fig.~3). 
The vanishing of the zero point energy occurs
 because of the particular value of
the zero point energy of the fermionic oscillator (\ref{HF}) which was forced
upon us by supersymmetry---if the zero point energies of the bosons
and fermions did not cancel, we could not have written $\hat{H}_B + \hat{H}_F$ in
the form (\ref{HSUSY1}). 
We will see in the next section that the cancellation of the zero point energies of 
the bosons and fermions due to supersymmetry is one of the central reasons to
 believe that supersymmetry may be present at energy scales of the order of 
the electroweak
 scale.

 The discussion in this section has had so far a rather formal character. 
 To illustrate the fact that supersymmetry exists in the 
 physical world, we note that  
 a   simple physical system has energy levels that precisely
 match those of the supersymmetric oscillator.
 One can  recognize on Fig.~3 the Landau levels of an electron moving in 
 a constant, homogeneous magnetic field $\cal{H}$, upon identifying 
 $\omega = |e|{\cal{H}}/(m_e c)$ (here $e, m_e$ are the electron
 charge and mass, and $c$ is the speed of light). It is well known\cite{LL} 
 that upon
 quantizing the classical Larmor orbits of an electron in a magnetic field, 
 only a discrete set of radii is allowed, 
 labeled by an integer radial quantum number $n = 0,1,2,...$.
 The quantum number $k = 0,1$, on the other hand,
  denotes the projection of the electron's spin
 on the direction of the magnetic field. Supersymmetry, therefore, 
  relates the  energy of an electron in the $n$-th Landau level with spin along the 
 magnetic field to the energy of an electron in the $n+1$-st level with 
 spin opposite  the magnetic field (see Fig.~4). We should also  
  note that this nonrelativistic
 supersymmetry is only approximate: the degeneracy of the electron
 Landau levels
  due to 
 supersymmetry is spoiled by many effects (e.g. the electron's anomalous magnetic 
 moment) that are not taken into account in the nonrelativistic Schr\" odinger
 approach.

\begin{figure}[ht]
\vspace*{13pt}
\centerline{\vbox{\hrule width 5cm height0.001pt}}
\vspace*{.1truein}	
\centerline{\psfig{file=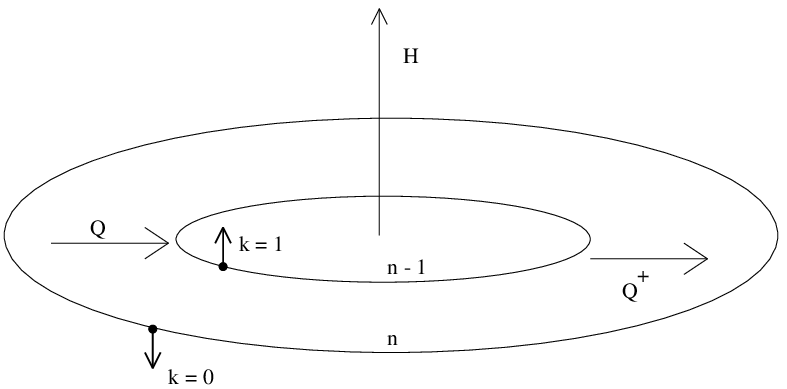}}
\centerline{\vbox{\hrule width 5cm height0.001pt}}
\vspace*{13pt}
\fcaption{The allowed orbits of an electron in a constant homogeneous magnetic
field $H$. The degeneracy of the two neighboring levels---the $n - 1$-st Landau
level with electron spin along ($k = 1$)
 the field and the $n$-th level with spin opposite ($k = 0$) the
field---is due to supersymmetry. The action of the supersymmetry 
generators $Q$ and $Q^\dagger$
is shown by the corresponding arrows.}
\end{figure} 

As emphasized in the beginning, the supersymmetric oscillator has
 most of the elements needed to 
generalize  supersymmetry to quantum field theory.
To this end, recall that the theory of photons---the quantized free Maxwell
field---can be considered as an infinite collection of simple harmonic oscillators.
Upon putting classical
 Maxwell electrodynamics in a large box and going to a  Fourier
representation, one observes that the field equations of the various Fourier
(i.e. momentum) modes decouple.
 The equation for  each momentum mode
  describes a  simple harmonic oscillator (with  
 momentum  playing the role of the coordinate). 
The quantum state of the momentum mode with momentum $\vec{k}$ and
 frequency\footnote{Hereafter we put $\hbar = c = 1$; for Maxwell
electrodynamics we take $m=0$.} $~~\omega (\vec{k},\lambda ) =
\sqrt{\vec{k}^2 + m^2}$ is described
by a single quantum number $n_{\omega (\vec{k},\lambda )}$---the number 
of photons with 
momentum $\vec{k}$ and polarization $\lambda$.

We can now repeat this quantization
procedure word for word starting with the supersymmetric oscillator. 
A supersymmetric field theory will be just the infinite sum of supersymmetric
oscillators. The spectrum of the theory is given by the sum of the spectra
of Fig.~3 for each momentum mode and polarization---in Fig.~3 we have to replace
the frequency $\omega$ with the frequency appropriate for the given momentum 
mode $\omega \rightarrow \omega (\vec{k},\lambda )$. Thus,  in 
supersymmetric Maxwell electrodynamics, as in the 
supersymmetric
oscillator, a single-photon state 
with a given momentum and polarization 
$(1_{\omega (\vec{k},\lambda )}, 0_{\omega (\vec{k},\lambda )})$
is, due to supersymmetry, 
 degenerate with the state 
 $(0_{\omega (\vec{k},\lambda )}, 1_{\omega (\vec{k},\lambda )})$ of
  a fermion of spin 1/2 
 in a state with given momentum and polarization. More generally, a
 state with $n$ photons of momentum $\vec{k}$ and polarization $\lambda$,
 $(n_{\omega (\vec{k},\lambda )}, 0_{\omega (\vec{k},\lambda )})$ 
 is degenerate with the state with
 $n-1$ photons and one   spin-1/2 fermion 
 $(n_{\omega(\vec{k},\lambda)} -1, 1_{\omega (\vec{k},\lambda )})$.
This new spin-1/2 
state is the supersymmetric partner or the photon, known  in the 
supersymmetric nomenclature as the {\it photino}. 
In supersymmetric Maxwell electrodynamics the photino 
 has the same mass and  quantum numbers as the photon, 
but half integer spin.

The construction of supersymmetric field theory can be generalized along the
above lines 
to include interactions, as well as particles of various spins. 
Doing this in any detail requires the introduction of new techniques
and  would take us far from the objective of this article (the interested
reader can consult refs.\cite{susyintro},\cite{martin}). 
However, it should be clear by now that, similar to the construction of 
the supersymmetric oscillator, one can ``supersymmetrize" the theory
that describes all known elementary particles: the standard model of 
elementary particle theory. Similar to the photon and photino, all
known elementary particles acquire supersymmetric partners, which have
the same quantum numbers (charges, masses, etc.) as the ordinary particles.
The superpartner of the electron is the spin-0 boson called the {\it selectron};
the quarks and the other 
leptons acquire spin-0 partners called {\it squarks} and {\it sleptons},
respectively;
the gluons---the spin-1/2
{\it gluinos}, the $W$ and $Z$ bosons---the spin-1/2 {\it winos} and {\it zinos}, etc. 
The whole
nomenclature of what has become known as the MSSM (the ``minimal supersymmetric
standard model") can be found in 
refs.~\cite{nilles}$^,$ \cite{haberkane}$^,$ \cite{martin}.
 
Once we include interactions, the supersymmetric  
partners of the ordinary quarks and leptons, the squarks and sleptons introduced 
above, will 
acquire interactions similar to those of the quarks and leptons (since 
now not only the harmonic terms in the Hamiltonian, but the nonlinear interaction
terms of the bosons and fermions will be related by supersymmetry). 
For example, the spin-0
  squarks and sleptons couple to the photon and the $Z$-boson in the same
  way as the quarks and leptons: the corresponding Feynman graphs 
  for electrons and selectrons are  shown on Fig.~5. 
  If supersymmetry was exact, 
  the squarks and sleptons would have the same mass
  as their quark and lepton superpartners. They
   would contribute (through the upper graph in Fig.~5)
  to the decay width of the $Z$ an amount similar to the contribution of the
  nonsupersymmetric particles. 
  The precision measurement at LEP
  determining the width of the $Z$, however, 
  could not have agreed so spectacularly with the 
  prediction of the nonsupersymmetric standard model (
  for $\Gamma_Z^{exp} = 2.4946 \pm .0027$ GeV, 
  while the theoretical prediction without 
  supersymmetric particles is $\Gamma_Z^{th} = 2.4972$ GeV).
  
\begin{figure}[ht]
\vspace*{13pt}
\centerline{\vbox{\hrule width 5cm height0.001pt}}
\vspace*{.1truein}	
\centerline{\psfig{file=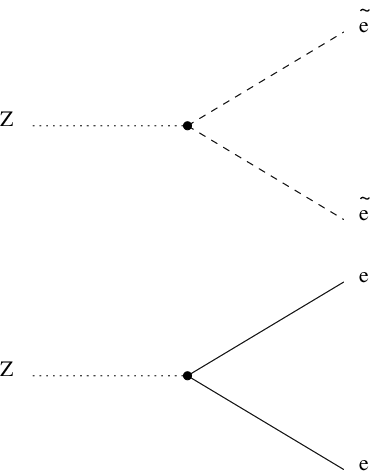}}
\centerline{\vbox{\hrule width 5cm height0.001pt}}
\vspace*{13pt}
\fcaption{Contributions of the electron supermulitplet 
to the decay width of 
the $Z$ boson in the supersymmetrized
standard model (the supersymmetric partners of 
the electrons are denoted by $\tilde{e}$).}
\end{figure} 

  The considerations from the previous paragraph 
  force us to conclude  that supersymmetry, when applied
  to the known elementary particles, can not be realized as in the supersymmetric
  oscillator---the spectrum of the known particles does not look anything
  like Fig.~3. 
From our everyday experience (and from many low-energy experiments)
we can conclude that the superpartners, if they exist, can not be degenerate in mass
with the ordinary particles (the strongest bound on their mass comes from
the LEP experiments mentioned above). 
  Does this mean that one must right away abandon the idea of
  supersymmetry? It is certainly possible that supersymmetry is a mathematical
  construction, remarkable for its inherent beauty but irrelevant for the physics
  of the elementary particles at energies near the electroweak scale. 
  It is also possible, however, that supersymmetry is just 
  around the corner, and a rich new spectrum of supersymmetric particles is
  waiting to be discovered just above the presently attainable energy. In the next
 section, we will review the theoretical arguments that make one believe that 
 this might be the case. 
 
 Before going on to that, recall the definition of supersymmetry that we gave
 at the beginning of this section. We hope to have elucidated the second part of
 the definition---that supersymmetry relates bosons to fermions---by means
 of our quantum mechanical example. However, 
  we did not give any  evidence
   that supersymmetry is a {\it space-time} symmetry. 
 To see that this is 
 the case, consider the defining relation (\ref{susyGENS})
  of the supersymmetry generators: $~\left\{ Q, Q^\dagger \right\} = \hat{H}$. 
 To generalize this relation to elementary particle theory, one has to promote
 it to a relativistically invariant relation. In a relativistically invariant
 theory, the Hamiltonian is  the zero component of the four-momentum vector, 
 while an 
 anticommuting object (such as the Dirac spinor field) has to carry spinor 
 indices under the Lorentz group. Heuristically, one expects to make
 the following replacements in (\ref{susyGENS}):
 \begin{eqnarray}
 \label{generalization}
 H  = P_0 &\rightarrow & P_{\mu} \nonumber \\
 Q &\rightarrow & Q_{\alpha} \\
 Q^\dagger &\rightarrow&  Q^\dagger_{\dot{\alpha}} \nonumber~.
 \end{eqnarray}
 In (\ref{generalization}), $P_\mu$ denotes the four-momentum vector, while
 $\alpha, \dot{\alpha}$ denote  spinor indices (under the $(0,{1\over 2}),
 ({1\over 2},0)$ representations of the Lorentz group, respectively). After
 these replacements, the anticommutation relation (\ref{susyGENS}) can be 
 written in a Lorentz covariant form
 \begin{equation}
 \label{n1susy}
 \left\{ Q_\alpha, ~Q^\dagger_{\dot{\alpha}} \right\} ~=~
 -~ 2 ~i ~ \sigma^\mu_{\alpha \dot{\alpha} } ~ P_\mu ~,
 \end{equation}
 known as (part of) the $N=1$ supersymmetric algebra in $3 + 1$ dimensions.
 (We stress again 
 that our goal here is to give a flavor of the subject and familiarize the reader
 with the main ideas;  
 we only note that 
 $\sigma^\mu_{\alpha \dot{\alpha}}$ are related to the $\gamma$ matrices,
 for details see e.g.\cite{susyintro}$^,$ \cite{martin}.)
 
 The relation (\ref{n1susy}) of the supersymmetry algebra
  implies
 that supersymmetry is a space-time symmetry---the anticommutator of two
 supersymmetry transformations (those generated by $Q$ and $Q^\dagger$) is
 a translation in space time generated by the momentum  $P_\mu$ 
 (eq.~(\ref{n1susy}) is often interpreted by 
  saying  that supersymmetry is a ``square root of
momentum"). The supersymmetry algebra (\ref{n1susy}) also implies
 that if we want to construct
a supersymmetric theory where supersymmetry is a gauge symmetry (i.e. the 
parameter of the transformation depends on the space-time point), 
we will necessarily have to gauge space-time translation (since, for consistency,
 both sides of  (\ref{n1susy}) will need to
represent local transformations). Since gauging space time translations is 
equivalent to constructing general relativity, gauging supersymmetry therefore
implies general relativity. This is a hint that there might be some deep 
connection between the structure of space time and supersymmetry.

\noindent

\section{Why is (broken) supersymmetry relevant in high-energy physics?}

From the discussion in the previous section, we learned that supersymmetry,
at least at energies below $\sim 100$ GeV, 
can not be linearly realized (i.e. lead to spectra like that on Fig.~3 ). 
Such a spectrum, with the superpartners 
degenerate in mass with the ordinary particles, is in blatant contradiction
with all known data.
All is not lost, however, and there is still hope that 
supersymmetry might be relevant in one form or another to physics
at energy scales accessible to the next generation of colliders.
This hope rests mainly on various theoretical ideas, which we  review in 
this section.

At the end of the previous section, we saw that supersymmetry is a space-time
symmetry: the generators of supersymmetry can in a loose sense be described
as square roots of the translation generators. We also saw that 
the anticommutation relation (\ref{n1susy}) of 
the $N=1$ supersymmetry
algebra means that local supersymmetry implies gravity (or rather, its 
supersymmetric extension known as supergravity,
which will come into play later on). Moreover, the only known consistent theory
of gravity, string theory, often 
implies the existence of space time supersymmetry (all known consistent vacua of
string theory have space time supersymmetry). 
The last statement needs some qualification: strictly speaking, superstring theory
would only imply the existence of supersymmetry at energy scales above or
 of order of the Planck
scale, $M_{Pl} = \sqrt{\hbar c/(8 \pi G_N)} = 
2.4 \cdot 10^{18}$ GeV ($G_N$ is Newton's gravitational constant). 
On the other hand, 
low energy experiments (and our everyday
experience) teach us  that the supersymmetric partners and the ordinary elementary 
particles are not degenerate in mass: the agreement of 
$\Gamma_Z$ between
theory and experiment implies that the masses of the superpartners of the 
ordinary quarks and leptons that couple to the $Z$-boson have to be at least
$m_Z/2$, so that they do not contribute to the width $\Gamma_Z$ of the $Z$-boson.
 Therefore, we are forced to conclude that supersymmetry has 
has to be ``lost" somewhere between $M_{Pl} \sim 10^{18}$ GeV 
and the electroweak scale, $M_{W,Z} \sim 10^2$ GeV.

String theory, in fact,
  provides only one of the  motivations for supersymmetry at or above
 the  electroweak scale. Another motivation, which historically preceded string 
theory, stems from considering  electroweak symmetry breaking and 
the ensuing 
hierarchy problem. In the following, we will discuss this motivation (which can
be called the ``naturalness" motivation).
  
  Recall that in the standard model, 
  the weak and electromagnetic interactions are
  described by a nonabelian gauge theory, with gauge group 
  $SU(2)\times U(1)_Y$.
  This nonabelian gauge symmetry is, however, not linearly 
  realized at energies below
  $100$ GeV (i.e. it does not give rise to only massless gauge bosons). 
  As is well known, the electroweak gauge group  $SU(2)\times U(1)_Y$ is broken
  down to $U(1)_{e.m.}$, describing Maxwell electrodynamics, at
  a scale of order $10^2$ GeV. The only massless gauge boson is thus  
   the photon. 
   The other three gauge bosons, which have been observed experimentally,
  of  $SU(2)\times U(1)_Y$---the $W^{\pm}$- and 
  the $Z$-bosons---obtain mass of the same order of magnitude 
  $\sim 100$ GeV. 
  The theory of the electromagnetic and weak interactions has 
  been subjected to numerous experimental tests and the agreement between theory
  and experiment is spectacular.
  
\begin{figure}[ht]
\vspace*{13pt}
\centerline{\vbox{\hrule width 5cm height0.001pt}}
\vspace*{.1truein}
 \centerline{\psfig{file=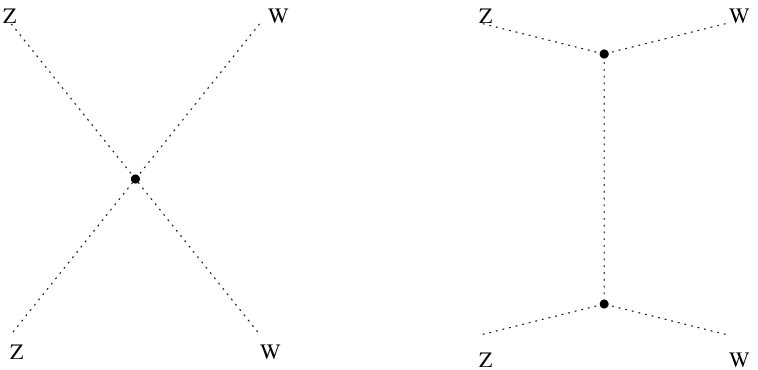}}
\centerline{\vbox{\hrule width 5cm height0.001pt}}
\vspace*{13pt}
\fcaption{The graphs contributing to the $Z Z\rightarrow W W$ scattering
process. The contribution of these graphs to the scattering amplitude violates
unitarity at energies $\sim 1000$ GeV.}
\end{figure}

  The only sector of the theory that has eluded experimental tests is the one 
  that is responsible for electroweak symmetry breaking and the generation
  of mass. That such a sector has to exist follows from considering the 
    the theory
  of massive spin-1 bosons, the $W^{\pm}$- and 
  the $Z$-bosons. It is well known that in
   theories of spin-1 massive particles 
the scattering amplitudes grow with energy. On Fig.~6, we have 
shown the Feynman diagrams responsible for the scattering process 
$Z Z \rightarrow W^+ W^-$. The amplitude of this scattering process 
grows with 
the center of mass energy $E$ like
\begin{equation}
\label{zzwwamplitude}
{\cal A}~(~Z~Z \rightarrow ~W^+~W^-~) ~\sim ~\left( {E\over 4 \pi v}\right)^2 ~,
\end{equation}
where $v$ is a scale of order  the mass of the $W$ boson.
The growth of the  amplitude with energy is disastrous---it 
implies that at energies of order $1000$ GeV some probabilities become larger
than one and unitarity is lost! This growth of the amplitude can be 
 stopped (and hence unitarity restored) 
by an exchange of, for example, a spin-0 boson---the Higgs boson---via
the graph of Fig.~7. Moreover, the introduction of the
Higgs boson sector is 
required for consistency of the theory (it supplies the longitudinal components
or the $W, Z$ bosons). Its expectation value breaks electroweak 
symmetry and generates the mass of the $W^\pm$- and $Z$-bosons.

\begin{figure}[ht]
\vspace*{13pt}
\centerline{\vbox{\hrule width 5cm height0.001pt}}
\vspace*{.1truein}
\centerline{ \psfig{file=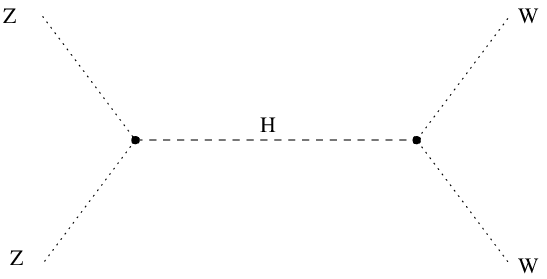}}
\centerline{\vbox{\hrule width 5cm height0.001pt}}
\vspace*{13pt}
\fcaption{Higgs boson contribution to $Z Z\rightarrow W W$ scattering.}
\end{figure}

So far, the Higgs boson (or more generally, the Higgs sector of the theory that
is responsible for electroweak symmetry breaking) has eluded experimental 
searches. Its mass is not fixed by the theory---it is an additional free
parameter in the Lagrangian. However, if the Higgs boson is elementary, both
its mass and expectation value have to be of the same order of magnitude,
i.e. one expects $m_H \sim 10^2 - 10^3$ GeV.

\begin{figure}[ht]
\vspace*{13pt}
\centerline{\vbox{\hrule width 5cm height0.001pt}}
\vspace*{.1truein}
 \centerline{\psfig{file=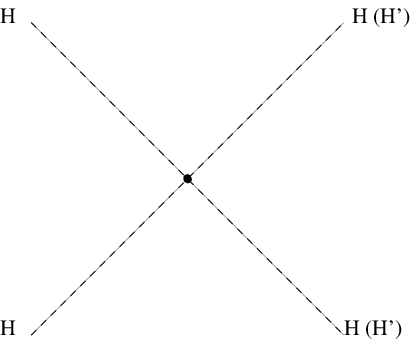}}
\centerline{\vbox{\hrule width 5cm height0.001pt}}
\vspace*{13pt}
\fcaption{The quartic scalar interaction vertex.}
\end{figure}

This is where we first encounter 
the {\it hierarchy problem}. This 
face of the hierarchy problem stems 
from the fact that  
theories of elementary scalar particles suffer severe naturalness problems. 
To elucidate, recall 
that elementary scalar fields can have pointlike four-particle interactions 
like the one on Fig.~8. Since the Higgs boson is a spin-0 particle, it can
have pointlike quartic interactions both with itself or with other,
possibly heavier, spin-0 particles 
(in the latter case, two of the lines in Fig.~8 would
represent the Higgs boson, while the other two would correspond to
 the other spin-0 particle, denoted by
$H^\prime$).
This interaction generates, via the vacuum fluctuation
 graph shown on Fig.~9, 
a contribution  to the Higgs mass
\begin{equation}
\label{deltamh}
  \delta m_H^2 ~\sim~ \left({\Lambda\over 4 \pi}\right)^2~.
  \end{equation}
   Here, $\Lambda$ denotes the ultraviolet cutoff,
or more physically, it is a scale of order of magnitude of the mass of
 the scalar particle running inside the loop.  
In most theoretical models that attempt to describe the  physics beyond the 
standard model such heavy scalar particles abound---this could be a 
particle responsible for the breaking of the Grand Unification symmetry
(and hence of mass $10^{15}$ GeV),
or it could  simply be one of the many Planck-mass particles left over from 
string compactification. The point here is that, once elementary 
scalar particles are allowed into the theory, there is no reason why 
some of them would not be as heavy as $10^{18}$ GeV. There is also absolutely  
 no symmetry 
reason why they would not couple to the Higgs via quartic pointlike 
interactions 
like the one on Fig.~8.

\begin{figure}[ht]
\vspace*{13pt}
\centerline{\vbox{\hrule width 5cm height0.001pt}}
\vspace*{.1truein}
 \centerline{\psfig{file=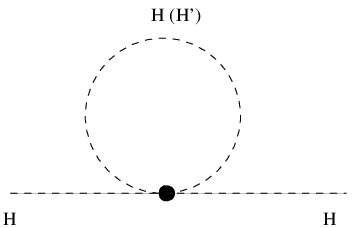}}
\centerline{\vbox{\hrule width 5cm height0.001pt}}
\vspace*{13pt}
\fcaption{Large contribution of radiative corrections to the Higgs mass.}
\end{figure}

Couplings like the one on Fig.~8 lie at the heart of the hierarchy problem,
for the correction to the Higgs mass from the graph on Fig.~9 is many orders
of magnitude larger than the value of the Higgs mass, the electroweak scale.
In order to achieve a Higgs mass of order $100$ GeV, we would have
 to fine tune both the bare value, $m_{H, 0}^2$ and $\delta m_H^2$, 
 so that
 \begin{equation}
 \label{finetuning}
 m_{H, 0}^2 ~+~ \delta m_H^2 ~=~m_H^2 ~\sim ~( 100 ~{\rm GeV} )^2~.
 \end{equation}
Since (\ref{deltamh})  $\delta m_H^2 \sim \Lambda^2 \sim
 ( 10^{15 - 18} {\rm GeV} )^2$, in order
to satisfy (\ref{finetuning}) we need to take $m_{H,0}$ 
of the same order of 
magnitude as $\delta m_H$ 
and achieve a cancellation between the bare value $m_{H,0}$
 and the correction $\delta m_H$
to order $10^{- 13} - 10^{-17}$! We would   
have to perform the same fine tuning 
not only at one loop, but in every order of perturbation theory.
This very unnatural situation leads to the 
 puzzle of why the electroweak scale---and
 hence the masses of the $W$, $Z$ bosons and the Fermi
constant---is so tiny compared to the Planck or Grand Unified scale. 
This puzzle is called the ``hierarchy problem."

The hierarchy problem, phrased above as the stability of the electroweak
scale against large radiative corrections, 
 arises because there is  no symmetry that 
protects scalar particle masses from receiving huge radiative corrections. 
Massless fermions, on the other hand, carry a conserved 
quantum number that protects them from obtaining mass
through radiative corrections: the projection of the spin onto
the direction of motion, known as chirality.
Since supersymmetry
interchanges bosons and fermions, in a theory that is supersymmetric, chirality 
will also protect the bosonic superpartners of the fermions from acquiring large
corrections to their masses. 
Thus supersymmetry in effect communicates chirality to the
fermions. 
More simply put, the graph of Fig.~9, the consideration of which
led us to the hierarchy problem, will have a supersymmetric partner in a theory 
with supersymmetry. The scalar running inside the loop would have a fermionic 
superpartner, which would contribute through the second graph on Fig.~10.
Since supersymmetry relates the couplings of the two graphs, and since 
the fermion contributes, due to the Pauli principle, with the opposite sign, the
leading contribution between the two graphs cancels, and only a 
small correction to the Higgs mass is left. We note that this cancellation
is akin to the cancellation of the zero point energies of the 
fermions and the bosons in the supersymmetric oscillator 
discussed in the previous section (see eqs.~\ref{HB}, \ref{HF}, \ref{HSUSY}).

\begin{figure}[ht]
\vspace*{13pt}
\centerline{\vbox{\hrule width 5cm height0.001pt}}
\vspace*{.1truein}
 \centerline{\psfig{file=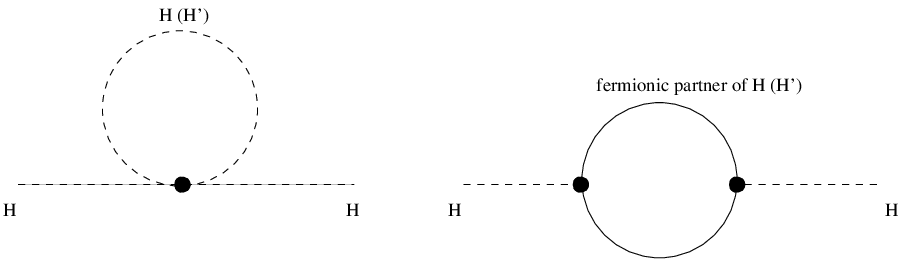}}
\centerline{\vbox{\hrule width 5cm height0.001pt}}
\vspace*{13pt}
\fcaption{Supersymmetric cancellations of large contributions
 of vacuum fluctuations to the Higgs mass, analogous to the 
 cancellation of the zero point energies of the bosons and fermions
  in the supersymmetric oscillator.}
\end{figure}

This is the way supersymmetry ensures the stability of 
the electroweak scale against large radiative corrections. Technically, 
the stability of the 
small electroweak scale against large radiative corrections
(i.e. its {\it naturalness}) is due to the cancellation of quadratic
divergences in supersymmetric theories. Due to these cancellations,   no
infinite number of fine tunings 
is required to keep the mass of the Higgs and the electroweak scale light.

We described the  way supersymmetry would solve the hierarchy problem
 in a theory in which supersymmetry is realized linearly,
and the spectrum is like that on Fig.~3. But the spectrum of elementary 
particles is not supersymmetric. So how are we going to take advantage of these
wonderful cancellations, brought by supersymmetry, to solve 
the hierarchy problem?
Remarkably (we will not be able to discuss details here), the cancellation 
of quadratic divergences (i.e. the absence of huge radiative corrections
to scalar masses) persists even is supersymmetry is not exact (i.e. the 
spectrum is not like that on Fig.~3), but is {\it softly} broken.
In the context of application to elementary particle physics, 
 the Lagrangian
that describes the theory of the softly broken supersymmetric
standard model consists of two parts:
\begin{equation}
\label{softsusy}
{\cal L}~=~ {\cal L}_{\it SUSY\ SM} ~+~{\cal L}_{soft}~.
\end{equation}
In (\ref{softsusy}),
 ${\cal L}_{\it SUSY\ SM}$ 
is the supersymmetric Lagrangian of the  standard model.
It describes the known elementary particles and their superpartners, 
contains no dimensionful (mass)
 parameters\footnote{We prefer to list the $\mu$-parameter,
which is formally supersymmetric, in the soft-breaking Lagrangian. For successful
phenomenology, $\mu \sim m_{soft} \sim m_W$ is required, 
and the only natural solution is to assume that the appearance of the
 $\mu$ term is related to supersymmetry breaking.},
  and can 
be constructed by the procedure outlined in the previous section for the
supersymmetric oscillator.  

In the absence of the second term in (\ref{softsusy}),
the spectrum of the elementary particles looks schematically like the one on
Fig.~3. 
The second term, the {\it soft-breaking Lagrangian}, ${\cal L}_{soft}$, 
is responsible for lifting the degeneracy of the spectrum and making 
(\ref{softsusy}) a Lagrangian consistent with experiment---it contains
the soft breaking terms
that include mass terms
 for the superpartners of all known elementary particles.
These mass terms, of order of 
magnitude $m_{soft}$, explain why the superpartners evade the bounds from LEP on
$\Gamma_Z$ and have eluded observation (so far). The mass of the Higgs boson 
 is also of order $m_{soft}$.
If supersymmetry is responsible for protecting the 
electroweak scale from large radiative corrections,
the natural value of the soft mass parameters is $m_{soft} \sim m_W \sim 
10^2-10^3$ GeV, for if all superpartners were much heavier than $m_W$, we would
have to explain another hierarchy problem: why $m_W \ll m_{soft}$.
The physics below the scale of
the soft breaking masses $m_{soft}$ is the physics of the  
nonsupersymmetric standard model---the Lagrangian (\ref{softsusy}) 
reduces to that of the 
nonsupersymmetric standard model, as the effect of the 
heavy supersymmetric particles is neglibigle at lower energy scales.

The Lagrangian  ${\cal L}= {\cal L}_{\it SUSY\ SM} + {\cal L}_{soft}$ is the main object of 
study of  supersymmetric phenomenology. 
The problem with using this Lagrangian is that $ {\cal L}_{soft}$ contains an
enormous number of parameters ($>100$). These are somewhat constrained by
low-energy measurements (see e.g. ref.~\cite{martin}, and 
the discussion of FCNC in Section 5), but still a large
degree of arbitrariness remains. How can we reduce (short of measuring the 
spectrum of superparticles) this large number of arbitrary 
parameters? 

In order to attempt to answer   this question, we note that 
the hierarchy problem has another face, in addition to the stability of the 
small electroweak scale from large radiative corrections.
Although in a theory
described by  ${\cal L}= {\cal L}_{\it SUSY\ SM} +{\cal L}_{soft}$,
with $m_{soft} \sim m_W$, the stability of the
electroweak scale against large radiative corrections is ensured, it is clear that
the Lagrangian (\ref{softsusy})
cannot be the ultimate Lagrangian of the universe, since it does not 
describe physics at arbitrarily small distances
(for example, it does not include
gravity). 
In fact, ${\cal L}$ still contains an enormous amount of
fine tuning, albeit only at tree level: the only scale that appears
 in ${\cal L}$ is of order  $m_W \sim m_{soft} \sim 10^{-16} M_{Planck}$. 
 From the point of view of a (more) fundamental theory (?string theory) that
describes the physics at short distances and  
 has only one fundamental scale ($M_{Planck}$),  
 the appearance and order of magnitude of 
 the parameters in ${\cal L}_{\it SUSY\ SM} +{\cal L}_{soft}$ is as big a puzzle as it
 was before the introduction of supersymmetry. A successful theory leading to 
 a solution of the
 hierarchy problem should provide a dynamical explanation 
 of the origin and magnitude of the soft parameters $m_{soft}$  
 in the low-energy Lagrangian. This means that 
 dynamics at some higher scale, $M_{\it SUSY}$, should be   
 responsible for the breaking of supersymmetry. The supersymmetry breaking
 dynamics should therefore generate the 
 soft breaking terms ${\cal L}_{soft}$
in the low-energy Lagrangian 
  with the right order of magnitude
 $ m_{soft} \sim m_W$. The dynamics that break supersymmetry
 will then naturally imply definite relations between the
different soft parameters, reducing thus the large arbitrariness.
 It is thus the dynamics of supersymmetry breaking that 
 is responsible for the smallness of the electroweak scale.  
Conversely, we are hopeful that measuring the soft parameters (i.e.  finding
 the superpartners and measuring their masses) will allow us to gain insight into
 the higher-scale dynamics responsible for the breaking of supersymmetry.
 
Before proceeding in the next section with a more technical (though still based
 on quantum-mechanical examples) discussion of the possible ways to
 break supersymmetry, let us summarize the main points of the previous 
 two sections:
 \begin{itemize}
 \item{Supersymmetry is a space-time symmetry that unifies bosons and fermions.}
 \item{Although not exact at low energies, supersymmetry can be relevant for
 high-energy physics, because:
 \begin{itemize}
\item{Consistent string theory vacua have space-time supersymmetry.}
\item{Supersymmetry can explain the smallness of the electroweak scale.}
\end{itemize}}
\item{Supersymmetry must be dynamically broken and the masses
of the superpartners (selectron, photino...) should result from that breaking. The
breaking of supersymmetry should also trigger electroweak symmetry breaking.}
\item{The idea of supersymmetry at the electroweak scale is falsifyable. 
If  superpartners are found, measuring their masses can
 give us a glimpse upon dynamics at higher  energy scales.}
\end{itemize}

It is therefore interesting to understand how supersymmetry can break.

\smallskip

Finally,  we would like to mention the possibility that 
  electroweak symmetry breaking is due to other, non-supersymmetric
dynamics without elementary scalar fields, such as technicolor. However, 
until the existence of  supersymmetry at the TeV scale
 is firmly disproved,  we should not give up on
theoretical investigations of, and experimental searches for,
  supersymmetry---we
would be missing a great opportunity to learn about physics 
at high-energy scales and
fundamental space-time symmetries.

\noindent

\section{How does supersymmetry breaking occur?}

In this section, we will discuss  some possible ways to break supersymmetry.
To this end, we will return to our discussion of supersymmetric quantum
mechanics. In order to allow for symmetry breaking, we will have
to include interactions and slightly generalize the discussion of Section~2. 

The quantum mechanical system we will use as an example is that of a  
spin-1/2 particle moving
on the line. The state of the spin-1/2 particle is described by a 
two-component wave function (a Pauli spinor),
$\Psi (x) = \left(  \begin{array}{c} \psi_1(x)    \\ 
                               \psi_2(x)      \end{array} \right)$.
                                The two components of $\Psi$ 
                                 describe the
wave functions of the particle with spin projections $+1/2$ and $-1/2$ 
respectively. 
The Hamiltonian of our spin-1/2 particle on the line is:
\begin{equation}
\label{hamiltoniansusyqm}
\hat{H} ~=~   - {1\over 2}~{d^2 \over d~ x^2}~+~{1\over 2}~
\left(~ d~W(x)\over d~x~\right)^2 ~+~{1\over 2}~
\sigma_3 ~\left(~ d^2~W(x)\over d~x^2~\right)~.
\end{equation}
Here and below $\sigma_{1,2,3}$ denote the Pauli matrices. 
That this Hamiltonian
is supersymmetric is reflected in the fact that it can be represented as the square
of a hermitean supersymmetry generator, just like (\ref{HSUSY1}).
The hermitean supersymmetry generator (this would be the
analog of $Q + Q^\dagger$ in the supersymmetric oscillator, 
see eq.~(\ref{HSUSY1})) 
is 
\begin{equation}
\label{susyqmgenerator}
{\cal Q} ~= ~-~{i \over \sqrt{2}} ~ \sigma_1 ~{d\over d~ x}~ + ~{1 \over \sqrt{2}}~
 \sigma_2~ {d ~W(x)\over d~x} ~.
\end{equation} 
Using (\ref{susyqmgenerator}) it is easy to check  explicitly\cite{wittendsb}
that ${\cal Q}^2$  yields the Hamiltonian (\ref{hamiltoniansusyqm}): 
$\hat{H} = {\cal Q}^2$.
The function $W(x)$ is called the {\it superpotential} and completely determines
the interactions.
 The various terms in the Hamiltonian (\ref{hamiltoniansusyqm}) 
have the following interpretation.
The first term is the kinetic energy of the particle; 
we put its mass equal to 1.
The second term is the potential energy---the particle
moves in a potential well $V(x) = (W^\prime)^2/2$. 
The third term describes the 
``spin-orbit interaction" $\sim \sigma_3 W^{\prime \prime}$.
The {\it superpotential}, $W(x)$, determines both the 
potential and spin-orbit terms in the
Hamiltonian (since they are related by supersymmetry)\footnote{We note 
that these 
formulas are quite similar to the ones that are obtained in $3+1$ dimensional
renormalizable 
supersymmetric field theory---all   interactions are derived by the derivatives
of a single function, the superpotential $W(x)$. 
In the field theory case, the  
``spin-orbit" term would correspond to the Yukawa interaction between the 
bosons and fermions in the supermultiplet.}. 

We are interested in the issue of supersymmetry breaking. In order to be able
to discuss it, we need to find an order parameter: a quantity that signals
whether supersymmetry is broken or not. In general, the 
spontaneous breaking of any symmetry means that although the dynamics is 
invariant under the symmetry, the ground state is not (a common example 
of spontaneous symmetry breaking is, e.g. the spontaneous
magnetization of a ferromagnet, which breaks the rotational symmetry). 
The noninvariance of the ground state $\vert 0 \rangle$ 
under supersymmetry transformations 
would mean that 
the supersymmetry generator ${\cal Q}$ does not annihilate the ground state, e.g.
${\cal Q}  \vert 0 \rangle \ne 0$. Consider now  the following chain of
equalities:
\begin{eqnarray}
\label{orderparameter}
E_0 ~&\equiv& ~\langle~ 0~\vert~\hat{H}~\vert ~0~ \rangle \nonumber \\
&=&~\langle ~0~\vert~ {\cal Q}~{\cal Q}~\vert ~0~ \rangle  \\
&=& ~\vert\vert ~{\cal Q}~  \vert~ 0~ \rangle ~\vert\vert^2 ~> ~0~, {\rm iff} ~
{\cal Q}~  \vert~ 0~ \rangle \ne 0 ~.\nonumber
\end{eqnarray}
Here, $E_0$ is the ground state energy and we used the fact that the Hamiltonian
is the square of ${\cal Q}$. The inequality in the last line is true whenever
supersymmetry is broken, i.e. ${\cal Q}  \vert 0 \rangle \ne 0$. We thus see
that the ground state energy of a supersymmetric system is positive if and only
if supersymmetry is broken, and zero if and only if
 supersymmetry is unbroken. The 
ground state energy is thus the order parameter for supersymmetry breaking. Hence
answering the question of whether supersymmetry is manifest or broken is equivalent to
finding whether the ground state energy vanishes or not.

At the classical level---ignoring the spin-orbit interaction and
the zero-point energies---this question is  
easy to answer. We only have to look at the graph of the 
potential energy $V(x)$. We have
shown three possibilites on Fig.~11. Fig.~11a shows a potential which is
everywhere positive. Thus, classically, the ground state energy is positive and
supersymmetry is broken. The potentials on Fig.~11b,c both allow for classical
states of zero energy, hence, classically, supersymmetry is unbroken. 

The classical 
approximation is of course not the whole story. 
It is natural to ask   whether quantum corrections can change the
classical answer. 
Fortunately, in   supersymmetric systems, it is often easy
to give the {\it exact} answers to questions about the ground state. 
The reason behind this power supersymmetry has can be traced to the fact
that the Hamiltonian is a total square (see eqn.~(\ref{HSUSY1})). Answering
the question of whether supersymmetry is broken is equivalent to 
finding whether the Hamiltonian has a normalizable eigenstate 
of zero energy. In a supersymmetric system, however, in
order to find the zero-eigenvalue state, we do not have to solve the second order
Schr\" odinger equation 
\begin{equation}
\label{secondorder}
\hat{H} ~\vert~ 0~ \rangle ~=~ 0~.
\end{equation}
 Since eq.~(\ref{orderparameter}) shows that $E_0 = 0$ if
 and only if ${\cal Q}~  \vert~ 0~ \rangle = 0$, it suffices, instead of (\ref{secondorder}), 
 to solve the first order equation 
\begin{equation}
\label{firstorder}
{\cal Q}  ~\vert~ 0 ~\rangle ~= ~\left(
~ -~{i\over \sqrt{2}}~\sigma_1  ~{d\over d~ x} ~ +~{1 \over \sqrt{2}}
~\sigma_2~ {d ~W(x)\over d~x} ~\right) ~\Psi_0 (x) ~=~0~.
\end{equation}
Compared to the second order equation (\ref{secondorder}), which, for a 
general superpotential can only be solved numerically, the first order equation
(\ref{firstorder}) can be  solved for an arbitrary superpotential
 $W(x)$. Using simple Pauli matrix algebra,
it is easy to check that 
\begin{equation}
\label{psizero}
\Psi_0 (x) = 
 e^{\sigma_3 ~W(x)} \left( \begin{array}{c}   c_1 \\  c_2 
 \end{array} \right)
= \left( \begin{array}{c} e^{W(x)}  c_1 \\
 e^{- W(x)} c_2 \end{array} \right)
\end{equation}
is the general solution of the zero-eigenvalue equation (\ref{firstorder}).
The solution for the ground state wave function $\Psi_0$ depends on two
integration constants $c_1, c_2$. From eq.~(\ref{psizero}) it follows
that $\Psi_0$ 
 is normalizable only in two cases:
 \begin{eqnarray}
 \label{normalizable}
c_1~&=&~ 0~,~ W(x) ~\rightarrow ~ + \infty ~~ {\rm as} ~ x ~
 \rightarrow~  \pm \infty~, ~{\rm or}
\nonumber \\ 
c_2 ~&=&~ 0~,~ W(x)~ \rightarrow ~- \infty~~ {\rm as} ~ x~ \rightarrow ~\pm \infty~. 
\end{eqnarray}
Thus a normalizable ground state of zero energy
exists only if the superpotential $W(x)$ is ``even at infinity", i.e. it
has the same limit  at both $x = \pm \infty$ 
($+$ or $- \infty$, since we assume here that
the spectrum is discrete). A smooth
function $W(x)$ 
 with this property will necessarily have an odd number of extrema (and  
its derivative 
 $W^\prime$---an odd number of zeros). But since $V(x) =  (W^\prime)^2/2$, this 
 means that the criterion for unbroken supersymmetry is that the 
 potential has an {\it odd} number of zeros.

\begin{figure}[ht]
\vspace*{13pt}
\centerline{\vbox{\hrule width 5cm height0.001pt}}
\vspace*{.1truein}
 \centerline{\psfig{file=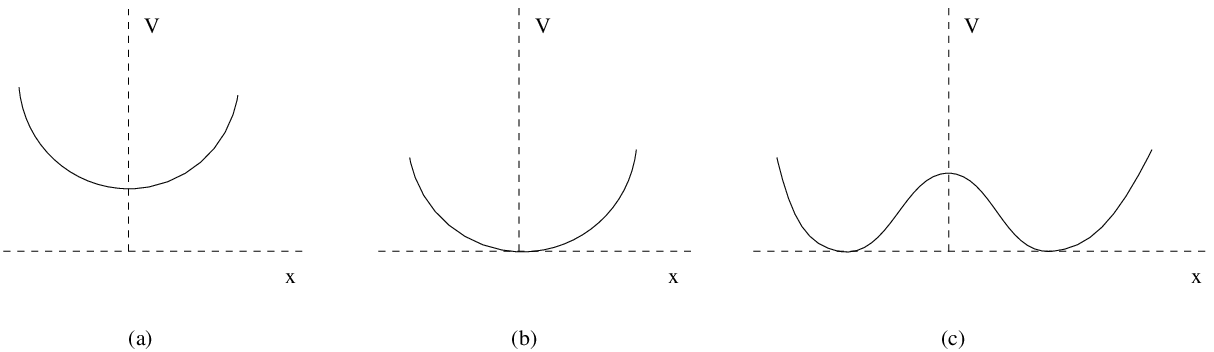}}
\centerline{\vbox{\hrule width 5cm height0.001pt}}
\vspace*{13pt}
\fcaption{Three possible potentials $V(x)$: (a) breaks
supersymmetry both at the classical and quantum level; (b) has unbroken supersymmetry
both classically and quantum-mechanically; (c) has manifest supersymmetry
at the classical level, but quantum-mechanical 
nonperturbative effects (tunneling)
break supersymmetry. All three possibilities have counterparts
 in quantum field theory.}
\end{figure}

We can now revisit the three potentials on Fig.~11 and 
find whether supersymmetry is broken
or not  at the exact quantum-mechanical level. The potential
on Fig.~11a has no zeros, hence according to our criterion from the
last paragraph, supersymmetry, being broken at the tree level, 
remains broken once quantum corrections are included. The potential on Fig.~11b
has one minimum, hence  supersymmetry remains unbroken in the quantum theory.
Finally, in the case of Fig.~11c, the potential has an even number of zeros. 
Therefore, 
even though supersymmetry is unbroken at the classical level, it is broken by
quantum effects.

It is the case depicted on Fig.~11c that will be of most interest for us. 
The  reason is that the breaking of supersymmetry in the 
supersymmetric system with a
double-well potential is due to nonperturbative effects---it occurs because
of tunneling between the two wells. 
We found earlier that in the classical approximation (and, even though we did not
show this, also in perturbation theory,
including the zero-point energy and the spin-orbit
interaction) the ground state energy vanishes and supersymmetry is unbroken.
The effect of tunneling can be evaluated in 
 the semiclassical approximation.
The WKB formula for the ground state energy splitting
 gives for the vacuum energy:
 \begin{eqnarray}
 \label{WKB}
 E_0^{WKB}~&=&~ \langle ~0 \vert~ \hat{H} ~\vert 0~ \rangle \nonumber \\
 &\sim &~ \hbar ~\omega~
  e^{ -{1\over \hbar} \int d x \sqrt{2 V(x)} } ~ \ll ~\hbar~ \omega~,
\end{eqnarray}
where $\omega$ is the  frequency of classical motion near the bottom of the 
well, and the integral is over the classically forbidden region of $x$.
Since, for appropriate parameters of the potential
(or,  in the formal semiclassical $\hbar \rightarrow 
0$ limit),  the 
tunneling probability
 is exponentially 
 suppressed, the scale of supersymmetry breaking---the ground state 
energy---is  much smaller than the characteristic 
frequency of motion inside the wells.  

The breaking of supersymmetry due to nonperturbative effects (similar
to the tunneling described here) has a counterpart in quantum field theory. The
typical expression one obtains for the scale of supersymmetry breaking,
$M_{\it SUSY}$, in field theory is very similar to the WKB formula (\ref{WKB}):
\begin{equation}
\label{msusyhierarchy}
M_{\it SUSY} ~\sim ~ e^{- O(1)~{8 \pi^2 \over g^2}}~M_{Planck} ~,
\end{equation}
where $g$ is  a  gauge coupling. The scale of supersymmetry breaking is 
thus exponentially suppressed compared to the Planck scale and can be many
orders of magnitude smaller. Since the electroweak scale is generated as a 
result of supersymmetry breaking, it  will also be much smaller than
the Planck scale (and, usually, smaller then $M_{\it SUSY}$).

There are  two main features of the discussion of our quantum-mechanical 
example that survive generalization to supersymmetric quantum field theory.
\begin{itemize}
\item{ Supersymmetry breaking is controlled by the  extrema
of the  superpotential.} 
\item{ Supersymmetry
breaking can occur {\it nonperturbatively} and 
generate {\it exponentially  small scales}, see eq.~(\ref{msusyhierarchy}).}
\end{itemize}

The quantum mechanical 
example we gave was intended to illustrate these two points. We  also
hope to have given a hint of 
the ``power of supersymmetry": in supersymmetric theories various
aspects of the dynamics, especially questions 
about the ground state, that are usually 
difficult to analyze (and are often 
intractable), can be understood exactly. 

The realization of this fact in $3+1$-dimensional 
supersymmetric field theory came after the work, in 1994, of Seiberg, and 
Seiberg and Witten\cite{SWreview}. It initiated what can be called
a ``supersymmetric revolution," which still continues. 
A tremendous progress has been made in 
understanding, even nonperturbatively, the low-energy 
dynamics of supersymmetric field theories. 
It was realized  that, in supersymmetric field theory,
  the superpotential (i.e. the appropriate generalization of $W(x)$ of 
  (\ref{hamiltoniansusyqm}))
 of  the lowest-energy excitations can be determined
exactly, including all nonperturbative effects. As we argued above, it
is  the superpotential that determines the phase structure of the theory. 
Thus, one gains a powerful tool to study the exact phase structure of 
supersymmetric field theories, and in particular, to answer the question:
which supersymmetric field theories break supersymmetry? 
While a simple criterion, as in our quantum-mechanical example (see discussion
after eq.~\ref{psizero}), is still absent, given a supersymmetric 
field theory, this question can be answered with certainty in most cases.
Indeed, since 1994, the list of theories for which the answer is known 
has grown dramatically. Many new theories and nonperturbative 
mechanisms of supersymmetry
breaking have been found (for   short
 reviews of this development and a list of references, 
 we refer the reader to \cite{wn}). 
 
 In this section, we illustrated the present  
 theoretical ideas of how supersymmetry can break nonperturbatively  
 and thereby generate small scales. The better understanding of the 
 physics of supersymmetry breaking, gained in the last several years,
 makes one hopeful  that   similar mechanisms 
 could be used to  explain the smallness of the electroweak
 scale and generate soft masses $m_{soft}$ for the superpartners
 of the  quarks, leptons, and gauge bosons.
 In the next section, we will describe the main ideas of how to 
use our understanding of supersymmetry breaking to construct 
phenomenological models that lead at low energies  to 
the  Lagrangian of the supersymmetric extension of the standard model
 (\ref{softsusy}).

\noindent

\section{How is SUSY breaking communicated?}

In this section, we describe the main current theoretical ideas
of how supersymmetry is broken and how the breaking manifests itself in the 
softly broken Lagrangian (\ref{softsusy}) of the supersymmetric standard model.

The first attempts to apply supersymmetry to elementary particle physics date
back to the early and mid 1970s. It was then quickly
 realized that the supersymmetrized
version of the standard model (i.e. $L_{\it SUSY\ SM}$ of 
(\ref{softsusy}), constructed
along the lines described in Section 1),
does not break supersymmetry. 
This meant that dynamics in addition to that of 
the standard model was required to
break supersymmetry. The existence of an additional sector of the theory was
thus postulated: the {\it supersymmetry breaking sector}. 

The supersymmetry breaking sector is usually a supersymmetric field theory, the
ground state of which breaks supersymmetry just as in our example
of supersymmetric quantum mechanics with  potentials of Figs.~11a and 11c.
If we want to use supersymmetry not only to stabilize the electroweak scale
against large radiative corrections, but also 
to explain its smallness compared to
the Planck scale,
the scenario of Fig.~11c is preferred  to
the one of Fig.~11a. 
This is because the theory with potential on Fig.~11a breaks 
supersymmetry already at the classical level.  Since  
 supersymmetry breaks at  tree level, the
scale of supersymmetry breaking is simply one of the parameters of the potential
and needs to be put in by hand (i.e. fine tuned). But
 a fine tuning of the ratio $m_W/M_{Pl} \sim 10^{-16}$ is precisely what we
 wanted to avoid. Hence
 the theory like that of Fig.~11c is more suitable to us: 
 the theory has unbroken supersymmetry at tree level, the scale 
 of supersymmetry
 breaking is generated dynamically  by nonperturbative effects and is 
 exponentially
 smaller than the fundamental scale of the theory, $M_{Pl}$. 
 
 To summarize, we expect 
 that the supersymmetry breaking sector is a theory that breaks supersymmetry 
 due to some nonperturbative effects. It is characterized by the scale of
 supersymmetry breaking $M_{\it SUSY}$ (which is related to the vacuum 
 energy density, as explained in the previous section). 
 In a theory of dynamically broken supersymmetry, 
 the scale of supersymmetry breaking is exponentially
 smaller\footnote{We note 
that obtaining an exponential suppression of $M_{\it SUSY}$ 
  requires that the factor in the exponent 
  is $\gg 1$, i.e. the  coupling 
  $g \le {\cal O}(1)$. This might look like another fine
 tuning problem.  
The gauge coupling $g$, however,
 requires much less adjustment than  fine tuning 
 the Higgs mass to 17 significant digits.} ~ than 
  the Planck scale (\ref{msusyhierarchy}): 
 $M_{\it SUSY} ~\sim ~ e^{- O(1)~{8 \pi^2 \over g^2}}~M_{Planck}$.

By now, we have discussed two of the  ingredients needed to construct 
a theory whose low-energy  effective 
Lagrangian is $L_{\it SUSY\ SM} + L_{soft}$: the supersymmetrized
standard model and the supersymmetry breaking sector. In order to generate the 
soft breaking parameters---the masses of the superpartners of the ordinary
standard model particles---these two sectors need to couple to each
other. The two main phenomenological frameworks for generating the soft 
masses ($L_{soft}$) are distinguished by the nature of the 
coupling between the supersymmetric standard model and the supersymmetry
breaking sector. The coupling between these two sectors has come
to be known as the {\it messenger interaction}. Presently there are two
main candidates to play the role
of messenger interactions---gravity and gauge interactions---and we discuss them
in turn.

\subsection{Supergravity mediated supersymmetry breaking.}

As explained in Section 2, supergravity arises naturally once supersymmetry
is promoted to a local symmetry.  Moreover, 
the low-energy effective description of string theory
is precisely the theory of supergravity.
 For our purposes, it will be enough to
state that in the theory of supergravity all 
elementary particles have their
superpartners, as described earlier. 
In addition, the massless spin-2 graviton has a supersymmetric,
massless 
spin-3/2 partner called the {\it gravitino}.  
Since gravity is a universal
interaction (it couples to the energy-momentum tensor), it  couples to 
both the supersymmetry breaking sector and the supersymmetric standard model.
Thus it is a natural candidate to play the role of a messenger interaction.
We will examine its consequences in what follows.

Once a supersymmetry-breaking theory is coupled to supergravity, 
an effect similar
to the Higgs effect takes place (the ``super-Higgs" effect). The 
important result of this effect is that the massless
gravitino obtains a mass $m_{3/2}$, which generally scales like
\begin{equation}
\label{gravitinomass}
m_{3/2} ~\sim~ {M_{\it SUSY}^2 \over M_{Pl}}~.
\end{equation}
Furthermore, 
the coupling of supergravity 
to the supersymmetric standard model also leads,
as a consequence of the classical equations of motion of 
supergravity,
 to the appearance of soft 
mass parameters for all scalar superpartners of the ordinary quarks, leptons, 
and gauge bosons (and for the Higgs). The natural order of magnitude of 
$m_{soft}$ is
\begin{equation}
\label{msoftsugra}
m_{soft}~\sim~m_{3/2} ~\sim~ {M_{\it SUSY}^2 \over M_{Pl}}~.
\end{equation}
This scaling can be obtained by dimensional reasoning. The scalar
masses for the superpartners enter as $m_{soft}^2$ in the Lagrangian.
Furthermore, since $m_{soft}^2$ arise as a consequence of the 
{\it classical} equations of motion of supergravity, they should be
proportional to the Newton constant $m_{soft}^2 \sim G_N \sim 1/M_{Pl}^2$. 
Hence the scaling of $m_{soft}$ with $M_{Pl}$ in eq.~(\ref{msoftsugra}).
The dependence on $M_{\it SUSY}$ can be then recovered by dimensional
analysis.

\begin{figure}[ht]
\vspace*{13pt}
\centerline{\vbox{\hrule width 5cm height0.001pt}}
\vspace*{.1truein}
\centerline{\psfig{file=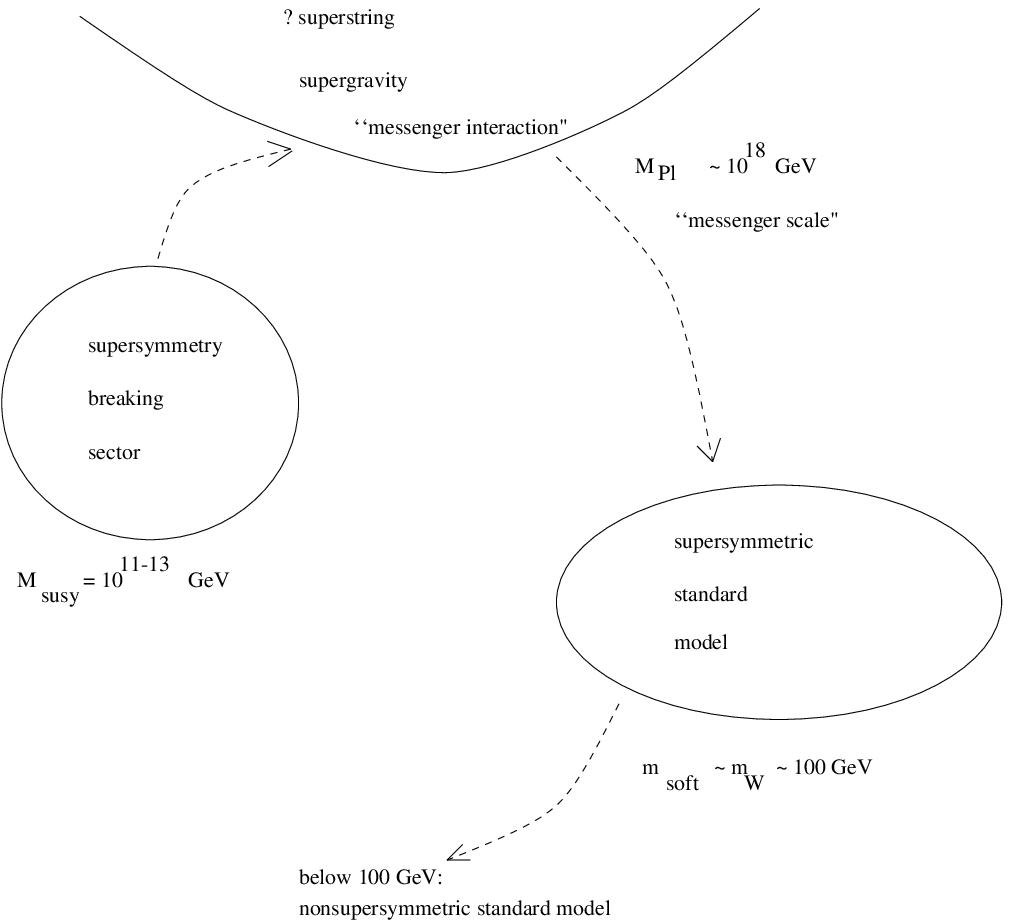}}
\centerline{\vbox{\hrule width 5cm height0.001pt}}
\vspace*{13pt}
\fcaption{Supergravity mediated (``hidden sector") supersymmetry breaking.}
\end{figure}

 In order to obtain the desired
$m_{soft} \sim m_W \sim 10^{2-3}$ GeV, 
taking into account $M_{Pl} \sim 10^{18}$ GeV,
eq.~(\ref{msoftsugra}) gives for the scale of supersymmetry breaking
$M_{\it SUSY} \sim 10^{11-13}$ GeV. Therefore, to explain the smallness of 
the electroweak
scale, the scale of supersymmetry breaking in supergravity mediated models
of supersymmetry breaking has to be of the order of 
an ``intermediate" scale---the geometric average of $m_W$ and $M_{Pl}$. 
The supergravity-mediated models are often called intermediate scale models,
or {\it hidden sector} models: since the scale of supersymmetry 
breaking is
so high, and since gravity is  the only interaction between the standard 
model and the
supersymmetry breaking sector, the supersymmetry breaking sector is ``hidden" 
from observation, up to energies of order $M_{Pl}$. On Fig.~12, we have 
schematically
shown  the structure and relevant scales in hidden sector models.

Supergravity mediated models have been in 
the center of study of supersymmetric
phenomenology since their appearance in the early 1980s. 
Among the reasons for
their popularity is the fact that supersymmetry breaking is communicated at
the classical (tree) level, and that
 the appearance of soft mass parameters is an 
automatic consequence of the coupling to gravity. 

However, the hidden sector
 models suffer from a severe drawback: they 
lack predictive power. The reason behind the lack of predictivity
 is that supergravity theories
are {\it effective} field theories,  only valid up to energy
scales of order $M_{Pl}$. The general Lagrangian of supergravity depends on
three arbitrary functions (to boot, some of them have indices). If we 
possessed detailed knowledge of  
the theory beyond the Planck scale, we could compute these
functions. However, in the absence of such detailed 
knowledge (the only candidate for such a theory, string theory,  as of now
does not allow matching to an effective supergravity 
theory with N=1  supersymmetry in $3+1$ dimensions) the Lagrangian remains 
rather arbitrary. 

As a consequence of this arbitrariness, the supergravity Lagrangian 
can lead to 
soft parameters that can be in conflict with experimental data. As mentioned
before, there exist  constraints from low-energy experiments on the 
soft parameters. One of these is the requirement of sufficient degeneracy of
the masses of the supersymmetric particles of the first two generations
that carry  the same
gauge quantum numbers.
This requirement 
 arises from the absence of flavor changing neutral currents contributions
to the $K^0 - \bar{K}^0$ mass difference; see ref.~\cite{martin}$^,$\cite{nilles}. 
The general supergravity
Lagrangian allows for nondegenerate scalar soft masses and is thus in
conflict with low-energy data. Various mechanisms have been proposed to 
remedy
this situation, but in the absence of understanding of the short-distance physics, 
it is hard to decide whether any of these can be operative.

To summarize, we believe that supergravity is an attractive mechanism of 
communicating supersymmetry breaking and generating the soft scalar masses
of the superpartners of the ordinary particles. Unfortunately, as
of now, it lacks predictive power. Recent developments in nonperturbative
string theory (string duality, M-theory) offer 
hope that this flaw may some day  be remedied.

\subsection{Gauge mediated supersymmetry breaking.}

 Historically, the first phenomenological models that 
 coupled the supersymmetry
 breaking sector to the supersymmetric standard model were the models where
 a gauge force plays the role of messenger interaction. They were proposed 
 also in the early 1980s, but before   the idea  of 
 supergravity as the messenger of supersymmetry breaking made its appearance.
 There are several reasons that they were abandoned: supergravity,
 where the soft parameters are generated at the classical level, won out with its
 simplicity, and, in addition,  at the time there was little understanding of the 
 nonperturbative dynamics of supersymmetric gauge theories.
 The gauge-mediated
 models were ressurrected after 1994, when  many aspects of 
supersymmetric
 nonperturbative gauge dynamics were better understood.
 
 It is natural to ask whether the standard model gauge interactions can
 play the role of the messenger of supersymmetry breaking. If this was the 
 case, the usual color and electroweak gauge interactions would have to also
  couple to the supersymmetry breaking sector.  The simplest possibility is
 that some of the ordinary quarks, leptons, Higgs, and their 
superpartners somehow
 participate in supersymmetry breaking. It was realized, however, that, at least
 in weakly coupled models, this direct coupling to the supersymmetry breaking
sector was  
disastrous: it led to spectra
 that were in conflict with experiment (some of the superpartners were lighter
 than the ordinary particles). The possibility that some of the known particles
 and their superpartners couple to the supersymmetry breaking sector 
remains if the models are 
 strongly coupled; however, at present, such models 
 lack predictive power.

It was then that the   {\it messenger quarks and leptons} were 
proposed as a remedy.
These are just like the ordinary quarks and leptons but are expected to 
be heavier (and hence undetectable at present). They are part of the 
so-called messenger sector, which is  coupled directly to
the supersymmetry breaking sector (see Fig.~13). As a result of supersymmetry
breaking, they obtain mass, of 
order $M_{mess}$---the {\it messenger scale}. Since supersymmetry is broken,
their spectrum is not exactly 
supersymmetric (i.e. unlike that of Fig.~3); the
degeneracy between the masses of the messenger quarks  and leptons and their
superpartners is lifted by supersymmetry breaking. Now, 
the messenger quarks and
leptons carry ordinary $SU(3)\times SU(2)\times U(1)$ quantum numbers. Therefore,
through quantum loop effects, they will also couple to the ordinary quarks and 
leptons (an example of such a radiative coupling that generates a soft mass
for a squark is given on Fig.~14). This is the way the squarks and sleptons 
``learn" about supersymmetry breaking and obtain soft masses. The magnitude
of the soft masses is easy to estimate:
\begin{equation}
\label{softmassgauge}
m_{soft} \sim {g^2\over 16 \pi^2} ~M_{mess} ~.
\end{equation}
The graph of Fig.~14 generates $m_{soft}^2$ at two loops, hence
$m_{soft}$ 
 is only proportional to a single loop factor (each loop contributes 
a factor of $g^2/(16 \pi^2)$).
In (\ref{softmassgauge}) 
$g$ is the relevant standard model gauge coupling.

\begin{figure}[ht]
\vspace*{13pt}
\centerline{\vbox{\hrule width 5cm height0.001pt}}
\vspace*{.1truein}
\centerline{\psfig{file=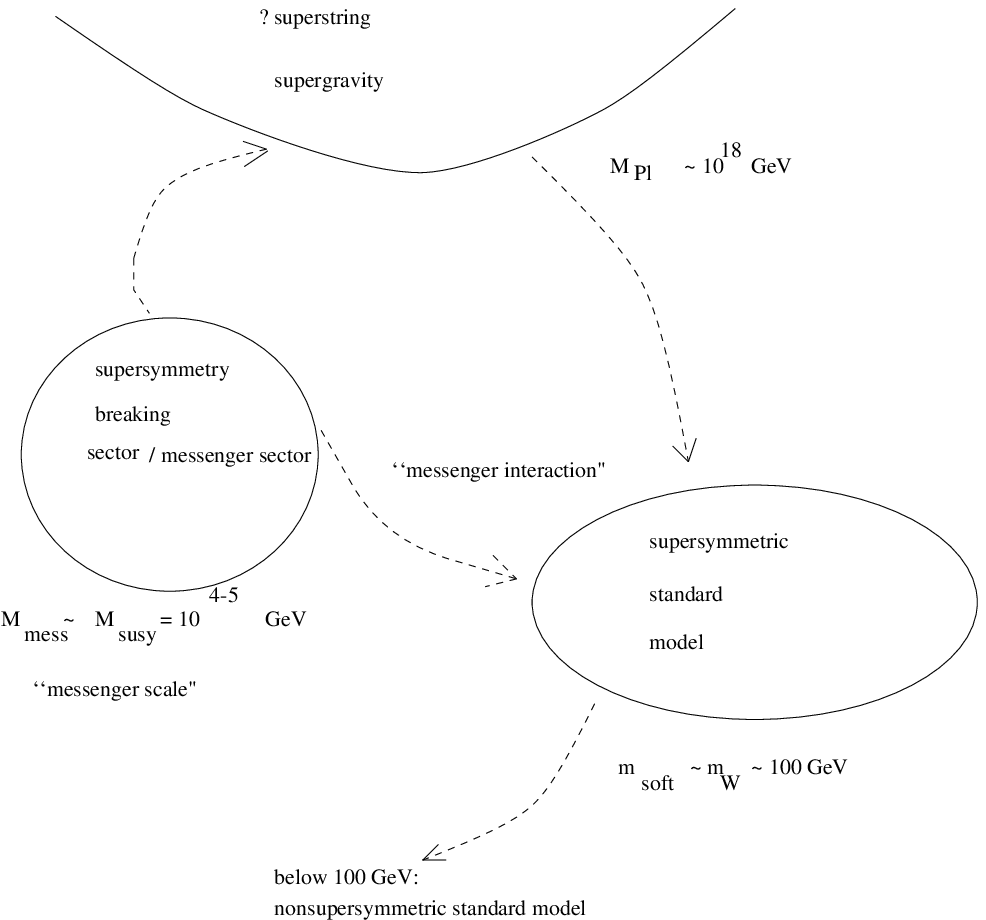}}
\centerline{\vbox{\hrule width 5cm height0.001pt}}
\vspace*{13pt}
\fcaption{Gauge mediated (``visible sector") supersymmetry breaking. 
We have assumed for simplicity that the scale of supersymmetry breaking,
$M_{\it SUSY}$ is of the same order of magnitude as the scale of the messenger
quarks and leptons, $M_{mess}$ (in general, 
these two scales need not be of the same order of magnitude). For a recent
review, see\cite{kolda}.}
\end{figure} 

Similar to what we did in the case of hidden sector models, we can estimate the
relevant scales. Demanding that $m_{soft} \sim m_W \sim 10^{2-3}$ GeV, we obtain
for the scale of the messengers: $M_{mess} \sim 10^{4-5}$ GeV (the supersymmetry
breaking scale $M_{\it SUSY}$ may or may not be of the same order of 
magnitude as the messenger scale; we crudely estimated the 
factor $g^2/(16 \pi^2) \simeq 100$). We note that the scale relevant for 
the communication of supersymmetry breaking in the gauge-mediated case is
many orders of magnitude smaller than that in supergravity. An important consequence
of eq.~(\ref{softmassgauge}) is that the soft scalar masses are proportional to
the gauge couplings, therefore   superpartners  with the same gauge
 quantum  numbers are automatically degenerate. 
Hence the flavor changing neutral 
 currents are naturally absent in gauge mediated models. This fact, and  their
 predictive power---because of their independence on the
short-distance physics---are the main arguments in favor of these models.

We should also mention the possibility of additional signatures
in gauge-mediated models, which do not
naturally arise in supergravity theories. To explain these, note that 
the supergravity effects are still there, but are not the leading ones
as far as the soft mass parameters are concerned (since their contributions
are suppressed by $1/M_{Pl}$). However, supergravity effects are
still the leading contribution to the gravitino mass: substituting $M_{\it SUSY}
\sim 10^{4-5}$ GeV in the formula for the gravitino mass (\ref{gravitinomass}),
we obtain $m_{3/2} \sim 10^{-10}-10^{-8}$ GeV---a mass in the eV range.
Such a light gravitino would lead to new signatures of supersymmetry (see 
\cite{kolda} and references therein).
 
\begin{figure}[ht]
\vspace*{13pt}
\centerline{\vbox{\hrule width 5cm height0.001pt}}
\vspace*{.1truein}
\centerline{\psfig{file=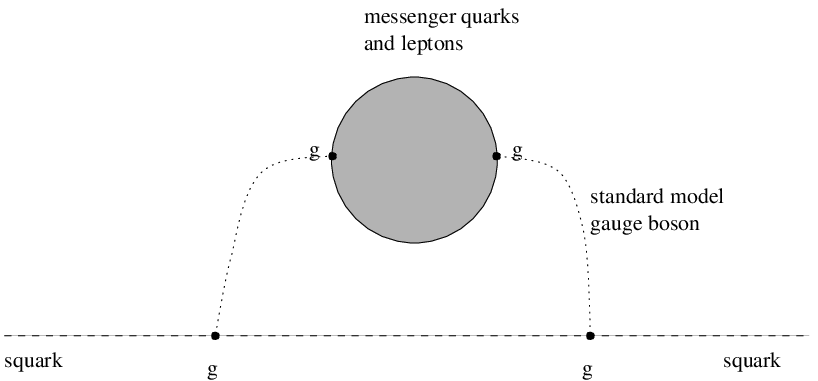}}
\centerline{\vbox{\hrule width 5cm height0.001pt}}
\vspace*{13pt}
\fcaption{An example of a two-loop contribution to the soft mass of the
squarks, slepton and Higgs in gauge mediated models. The usual quarks,
leptons and their superpartners couple to the messenger quarks and leptons
(and to the supersymmetry breaking sector) only through 
standard model gauge interactions,
hence  two loop graphs represent the leading contribution.}
\end{figure} 

 The explicit construction of models that realize the ideas discussed above,
 and are phenomenologically acceptable and elegant (admittedly the last criterion
 is rather subjective)  has not been  so successful,
  however. Nothing resembling a ``standard model" of gauge 
 mediated supersymmetry breaking has emerged quite yet.  
 It is perhaps fair to say that the lack of predictive power of
 supergravity has been replaced with a multitude of models, which are
  successful in  varying degrees (for a recent review and a list 
  of references see \cite{kolda}).
 The study of the possible experimental consequences of 
gauge-mediated models,
 perhaps in a model-independent way, is still a matter of importance. Once
 the superpartners are found and their spectrum measured
 it can help pin down the right mechanism of supersymmetry breaking.

\noindent

\section{Summary.}

Here we summarize the main points discussed in this review:
\begin{itemize}
\item{Supersymmetry is a 
space-time symmetry that interchanges bosons with fermions.}
\item{Even though it is not exact, supersymmetry can be relevant for elementary particle
physics, since:
\begin{itemize}
\item{Consistent string theory vacua have space time supersymmetry.}
\item{Dynamically broken supersymmetry can explain the smallness of the electroweak
scale, $m_W \sim 10^{-16} M_{Pl}$, and the masses of the superpartners. 
The idea of
supersymmetry breaking at the electroweak scale is falsifyable
in the near future.}
\end{itemize}}
\item{Recent advances in the study of 
supersymmetric quantum field theory allow us to
exactly study aspects of the supersymmetry breaking dynamics. 
Many new models and mechanisms
of supersymmetry breaking have been found in the last several years. }
\item{These advances
allow us to gain better 
insight into the phenomenological possibilities for 
supersymmetry breaking, 
make predictions, and rule out models.}
\end{itemize}

On the purely theoretical side, 
recent developments in string theory---string duality, D-branes,
and M-theory---offer hope of more complete understanding of the
dynamics of supersymmetry  breaking, in both  supergravity 
and gauge mediated frameworks. Perhaps some progress can be made
towards solving some of the 
many outstanding problems of supersymmetric phenomenology
(the stabilization of the dilaton,
the cosmological constant problem, etc.).

As emphasized in the text, there is no ``standard model" of supersymmetry breaking.
There exist a variety of models, none of which is entirely phenomenologically 
satisfactory, or has a particular
aesthetic appeal. In view of this, it appears 
that experimental input\cite{bagger}
 is important for
the further development of the field. 
One's  hope is that a discovery of supersymmetry
 will yield important hints towards the true mechanism of supersymmetry
breaking. 

\noindent

\section{Acknowledgments}

It is a pleasure to thank W. Skiba for comments on the manuscript.
This work was supported by DOE contract no. DOE-FG03-97ER40506.

\end{document}